\documentclass[useAMS,twocolumn]{emulateapj}
\usepackage[latin2]{inputenc}                                                                                                  
\def\apjl{{ApJ}}                
\def\apjs{{ApJS}}               
\def\ao{{Appl.~Opt.}}           
\def\aap{{A\&A}}                

\usepackage{graphicx}

\newcommand{\mpc}{\rm {h^{-1}Mpc }}
\newcommand{\ud}{\textrm{d}}

\begin{document}

\title{ISW Imprint of Superstructures on Linear Scales}

\author{P\'eter P\'apai}
\affil{Institute for Astronomy, 
University of Hawaii at Manoa}
\affil{Department of Physics and Astronomy, 
University of Hawaii at Manoa}

\author{Istv\'an Szapudi}
\affil{Institute for Astronomy, 
University of Hawaii at Manoa}

\author{Benjamin R. Granett}
\affil{INAF Osservatorio Astronomico di Brera}

\begin{abstract}
 
We build a model for the density and integrated Sachs-Wolfe (ISW) profile of supervoid and supercluster structures. Our model assumes that fluctuations evolve linearly from an initial Gaussian random field. We find these assumptions capable of describing N-body simulations and simulated ISW maps remarkably well on large scales. We construct an ISW map based on locations of superstructures identified previously in the SDSS Luminous Red Galaxy sample. A matched filter analysis of the cosmic microwave background confirms a signal at the $3.2-\sigma$ confidence level and estimates the radius of the underlying structures to be $55 \pm 28 \mpc$. The amplitude of the signal, however, is $2-\sigma$ higher than $\Lambda$CDM predictions. 

\end{abstract}

\keywords{large-scale structure of universe --- cosmic microwave background --- methods: statistical}

\section{Introduction}

Since the first sky maps of the Wilkinson Microwave Anisotropy Probe (WMAP) were published, there have been claims for the existence of circular features (spots and rings) in the cosmic microwave background (CMB) \citep{Cruz2005,Granett2008}. Their origin and statistical significance are still debated. Sources, such as foreground contamination, integrated Sachs-Wolfe (\citealp{SW1967}; ISW) effect and the more exotic conformal cyclic cosmology or cosmic texture have been considered viable candidates to explain circular CMB anomalies \citep{Rudnick2007,Inoue2010,Cruz2007,bs}. A notable feature is the cold spot \citep{Cruz2005}, which has a mean temperature of $-70\mu K$ in a $5^{\circ}$-radius aperture.  Additionally, $10\mu K$ hot and cold spots have been identified on $4^{\circ}$ scales associated with super structures \citep{Granett2008}.

In this paper we focus on the integrated Sachs-Wolfe effect and investigate whether it is possible that large spherical fluctuations (supervoids and superclusters) in the dark matter field produce the aforementioned features in the CMB. The large-scale ISW effect is expected in a universe with accelerated cosmic expansion arising either from a dark energy term in flat cosmological models, or from spatial curvature.  The effect is sensitive to both the expansion history and the rate of structure formation and provides constraints on alternative cosmological models (e.g. \citealp{Giannantonio2008b}) as well as initial non-Gaussianity (e.g. \citealp{Afshordi2008}). Cross correlating a galaxy catalog and a CMB temperature map is the standard way of studying the ISW signal and it has an extensive literature \citep{Scranton2003,Afshordi2004,Padmanabhan2005,Raccanelli2008,Giannantonio2008,Ho2008,Sawangwit2010}. Our work is similar considering what we measure is related to the correlation between the dark matter and the CMB. However, we only focus on parts of the sky where the signal is expected to be large, regions corresponding to supervoids and superclusters. Since reports \citep{Cruz2005,Granett2008} indicate that the scale of these regions is beyond nonlinear scales, our model for their average density profile and ISW imprint is derived from the statistics of the linearly evolving primordial Gaussian density field \citep{Papai2010}. This is in contrast with \cite{Inoue2010}, who assumed a top hat density profile, which is the asymptotic final state of a void with steep initial density profile \citep{Sheth2004}. 

We use the publicly available Hubble Volume Simulation of the Virgo Supercomputing Consortium (\citealp{Colberg2000}; HVS) to test the Gaussian model. This is the N-body simulation with the largest volume, which is relevant considering that the gravitational potential is correlated more strongly than the density. We simulate ISW maps by ray tracing through the potential and calculating the linear part of the ISW effect. These are compared to partial ISW maps generated from sets of spherical regions based on our model. 

After gaining some confidence by studying simulations we create an ISW map from real data. We select locations on the sky based on \cite{Granett2008}. They compiled a list of supervoids and superclusters found in the Sloan Digital Sky Survey (SDSS) Luminous Red Galaxy (LRG) sample. The list can be found in \cite{supplement}. We build an ISW map by placing the theoretical profiles to the given R.A., decl. coordinates. This map is fitted to a WMAP temperature map. 
 
The structure of the paper is the following: in Section \ref{PinS} we measure the expected density and ISW profile of spherical dark matter fluctuations in N-body simulations; in Section \ref{Match} we apply the matched filter technique to detect the signature of superstructures in the CMB; 
in Section \ref{discussion} we discuss our results and views of the relationship between the linear ISW effect and circular features on the CMB.  

\section{Profiles in Simulations}
\label{PinS}

In this section we calculate the full expected density profiles and ISW imprints of spherical overdensities and underdensities based on \cite{Papai2010} and compare them with measurements from the HVS.

\subsection{Density Profiles of Superstructures in N-body Simulations}
\label{densPr}

According to \cite{Papai2010}, if the average density contrast in a Gaussian random field inside a sphere of radius $R$ is given, we expect the density contrast at radius $r$ from its center to be:
\begin{eqnarray}
&\big<\delta(r)\big>_{\delta_{in}(R)} = \frac{\big<\delta(r) \delta_{in}(R)\big>}{\big<\delta_{in}^2(R)\big>}\delta_{in}(R),& \label{aveDelta}
\end{eqnarray} 
where $\big<...\big>$ stands for ensemble averaging, $\delta(r)$ is the density contrast at radius $r$ and $\delta_{in}(R)$ is the average of $\delta$ inside radius $R$. In the Appendix of \cite{Dekel1981} the author derived a similar formula for a single, point-like location with given density, which is the $R\rightarrow 0$ limit. Equation (\ref{aveDelta}) is essentially the two-point function multiplied by a normalization constant. A more general calculation for the density profile around local maxima in a Gaussian random field was carried out by \cite{BBKS1986}

In order to measure $\big<\delta(r)\big>_{\delta_{in}(R)}$, one needs to select regions with $\delta_{in}(R)$ and calculate the average over these. However, according to Equation (\ref{aveDelta}), $\frac{\delta(r)}{\delta_{in}(R)}$ gives an unbiased estimator of $\frac{\big<\delta(r) \delta_{in}(R)\big>}{\big<\delta_{in}^2(R)\big>}$, the shape of the profile:
\begin{eqnarray}
\big<\frac{\delta}{\delta_{in}}\big> = & \int{\ud \delta \ud \delta_{in} \frac{\delta}{\delta_{in}}P(\delta|\delta_{in})P(\delta_{in}) } \nonumber \\
 = & \frac{\big<\delta \delta_{in}\big>}{\big<\delta_{in}^2\big>} \int{ \ud \delta_{in} P(\delta_{in}) }  = \frac{\big<\delta \delta_{in}\big>}{\big<\delta_{in}^2\big>}, \label{normDelta}
\end{eqnarray}
where we shorten $\delta(r)$ and $\delta_{in}(R)$ as $\delta$ and $\delta_{in}$. We also use the fact that by definition $\big<\delta(r)\big>_{\delta_{in}(R)} = \int{\ud \delta \delta P(\delta|\delta_{in}) }$ and $\frac{\big<\delta \delta_{in}\big>}{\big<\delta_{in}^2\big>}$ is independent of $\delta_{in}$. The argument above shows that it is needless to search for spheres with a certain $\delta_{in}(R)$ to verify Equation (\ref{aveDelta}). In the rest of this Subsection we deal with the integral of $\delta(r)$ instead of $\delta(r)$, because the former is measurable directly:
\begin{eqnarray}
& M(r) = 4\pi\int_0^r \ud r r^2\delta(r). & \label{mass}
\end{eqnarray}

\begin{figure}	
      \begin{center}
        \includegraphics[scale=0.45]{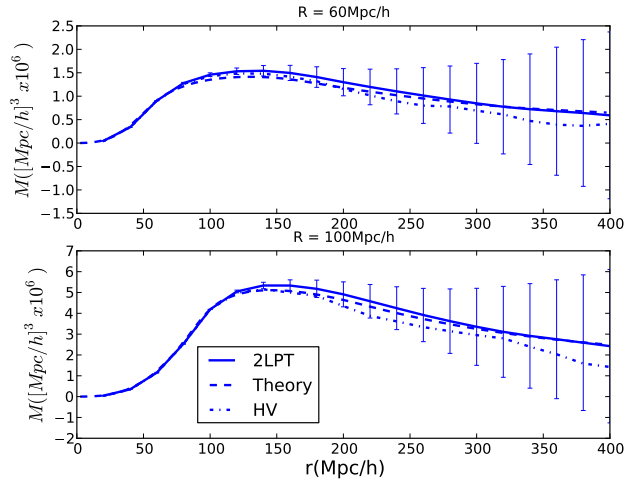}
      \end{center}
        \caption{The integral of the spherically averaged density profile measured in the HVS and 100 2LPT simulations. Linear theory given by Equation (\ref{normDelta}) appears to be a good approximation. \label{HVsig2}}
\end{figure}

The linear size of the HVS is $3000\mpc$. We chose to place spheres at $25^3$ evenly spaced coordinates. We counted galaxies in concentric spheres with radii of $10\mpc, 20\mpc, ..., 160\mpc $ and subtracted the average to get $M(r)$. We normalized this as $\frac{M(r)}{M(R)}\frac{4\pi}{3}R^3$ and calculated the average of the $25^3$ instances to get an estimate of the integral of Equation (\ref{normDelta}). As $R$ changes, technically, by weighting $M(r)$ differently, one can get the profile of linear superstructures of different sizes. A nuisance, we have to deal with, is that the estimator given by Equation (\ref{normDelta}) is unstable because $\delta_{in}$ can be arbitrarily small. The ratio of two Gaussian variables with zero means has a Lorentzian distribution, which has an undefined variance. This problem can be avoided if cases, when $|\delta_{in}|$ is under a threshold, are ignored. This constraint leaves the estimator unchanged as it can be seen from Equation (\ref{normDelta}). We chose the threshold based on the variance of $\delta_{in}$ in the particular catalog, e. q. $2\sqrt{\big<\delta_{in}^{2}\big>}$. The result is not sensitive to the exact choice but the right choice can decrease the variance of the estimator substantially. In Figure \ref{HVsig2} we plotted the measured profiles and their theoretical counterparts for $R = 60\mpc$ on the upper panel and for $R = 100\mpc$ on the lower panel. We obtained the error bars by repeating the measurement on one hundred mock catalogs generated with a second order Lagrangian (2LPT) code \citep{Crocce2006}. The mocks were set up to have the same cosmological parameters, size and density as the HV simulation. To calculate the theoretical profiles of Equation (\ref{aveDelta}) we used a matter power spectrum calculated via CAMB \citep{lewis}. The measured profiles, both in 2LPT and the HVS, appear to be in good agreement with theory within the uncertainty up to $400\mpc$, the largest radius in our measurement.

\subsection{The ISW imprints of superstructures} 

In \cite{Papai2010} the ISW effect in the direction of the center of a superstructure was calculated. This was done simply by calculating the spherically symmetric potential of the superstructure, then computing the following integral along the path of the CMB photon \citep{SW1967}:
\begin{eqnarray}
& \Phi (r) = -\frac{3\Omega_m}{8\pi}\big( \frac{H_0}{c} \big)^2 \int_{r}^{\infty} \frac{M(\tilde{r})}{\tilde{r}^2}\ud \tilde{r}\label{pot}, & \\
& \frac{\Delta T}{T} = -\frac{2}{c^2}\int \ud \tau \frac{\partial \Phi (r(\tau),\tau) }{\partial \tau}, &
\label{eq:isw}
\end{eqnarray}
where $\tau$ denotes conformal time. Similarly, this calculation can be carried out for any direction easily in order to predict the full ISW profile. In linear theory the potential at a a particular redshift is just $\Phi(r,z) = \Phi(r,z=0)\frac{D(z)}{a(z)}$. We leave the study of nonlinear corrections for future work. Since these integrals are linear, the linear ISW signal is predicted to be proportional to the average density of the superstructure, $\delta_{in}$. In this paper we place the centers of the superstructures at $z=0.52$, which is the median redshift of the galaxy catalog we use in Section \ref{Match}.

\subsection{Fluctuations in the Potential on the Largest Scales and the ISW Profile}

To test the ISW profile we traced rays through the HV simulation along the $z$-axis. We calculated $\delta$ on an $800^3$ grid and calculated $\Phi$ at $z=0$ by using the simple formula:
\begin{eqnarray}
 & \Phi(k) = -\frac{3\Omega_m}{2}\big( \frac{H_0}{c} \big)^2\frac{\delta(k)}{k^2}. &
\end{eqnarray}
Since we aimed to analyze imprints of superstructures at redshift $0.52$, we defined the boundaries of the integral of Equation (\ref{eq:isw}) in a way that put the $25^3$ preselected locations (see Subsection \ref{densPr}) at this redshift. We used periodic boundary conditions and we integrated up to $3000\mpc$ from starting points ensuring that $25^2$ of the locations were at redshift $0.52$ each time. This yielded 25 ISW images of the HVS in the $x-y$ plane. Each of these images contains imprints of $25^2$ superstructures. After this, one can follow the procedure discussed for density profiles in Subsection \ref{densPr}. If the ISW profile of a superstructure is normalized by the average density of the superstructure, the result will be an estimator of the shape of the ISW profile:
\begin{eqnarray}
& \big<\frac{\Delta T(r_{2D})}{\delta_{in}(R)}\big> = f(r_{2D},R). & \label{tprof}
\end{eqnarray}              
The arguments are $r_{2D}$, the radius in the $x-y$ plane and $R$, the radius in which we know the galaxy count in the corresponding volume. 

\begin{figure}	
     \begin{center}
        \includegraphics[scale=0.35]{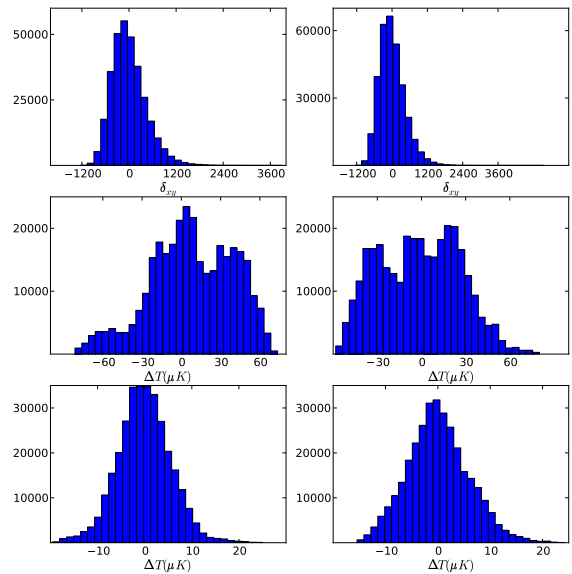}
     \end{center}
        \caption{Each row shows histograms of a certain projection of density along the $z$-axis. A simple integral of $\int \delta(x,y,z)\ud z$ is presented in the first row, an ISW map in the second and its filtered version in the third. (See text for details.) The $x-y$ plane is halved to test the cosmic variance, hence the two columns. As a result of large fluctuations in the low $k$-modes in the potential, the ISW projections show a significant cosmic variance (second row) which can be reduced by filtering (third row). \label{hist}}
\end{figure}

Here, we would like to remind the reader that $\big<...\big>$ refers to ensemble averaging. For density, averaging over a relatively large volume proved to be a sufficient substitute in Subsection \ref{densPr}. Despite the fact that the cosmic variance is large on the largest scales, these modes contribute only a little to the density distribution since their amplitudes are small. (The power spectrum goes to zero with $k$.) For the potential, this picture changes due to the $1/k^2$ factor. As a result, the ISW effect has a large cosmic variance regardless of the large size of the simulation. To demonstrate this, we projected the density to the $x-y$ plane. We split this projection in the middle of the $x$-axis to create two $1500$ X $3000 Mpc^2h^{-2}$ areas. In Figure \ref{hist} in the top row we plotted the histograms of these. We did the same with one of the 25 ISW images, which is essentially the projection of the potential. Those histograms are shown in the middle row. The cosmic variance is significantly larger for the ISW map. To demonstrate that the difference is due to modes on the largest scales we removed the low $k$-modes from the potential. After experimenting we found that removing modes with $|k|<2\pi\frac{5}{L}$, where $L=3000\mpc$ for the HVS, gave a visually compelling result (Figure \ref{hist}, bottom row). After applying this high-pass filter to a $2$-dimensional projection, the effect of the cosmic variance on the modes with lowest wavenumber will be around $18\%$.   

\begin{figure}	
        \begin{center}
        \includegraphics[scale=0.45]{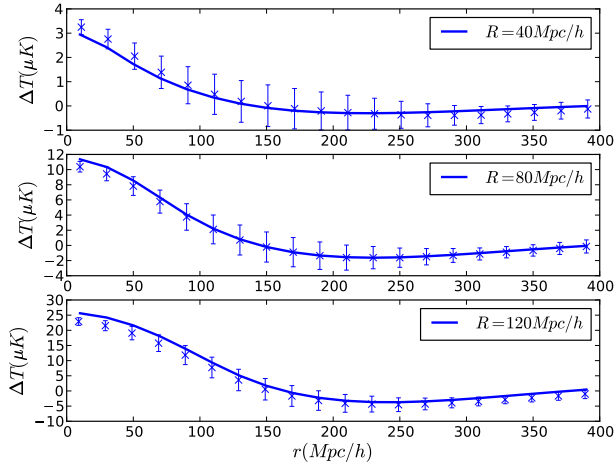}
        \end{center}
        \caption{The spherically averaged ISW profile from the HVS. This is defined by Equation (\ref{tprof}). Fourier-modes from the density were filtered out above $\lambda = 0.2L$, where L is the linear size of the HVS. The linear approximation is calculated as described in the text. See Equations (\ref{normDelta})$-$(\ref{eq:isw}). \label{iswprofHV}}
\end{figure}

Having removed the low $k$-modes from both the density and the potential we computed the average in Equation (\ref{tprof}). As before, we omitted cases when the denominator was too close to zero. (See Subsection \ref{densPr}.) The results for $R=60\mpc$ and $R=100\mpc$ are shown in Figure \ref{iswprofHV}. The error bars show the uncertainty calculated from the HVS itself due to the long CPU time required to calculate the ISW images. Predictions of the linear model are calculated by removing the low $k$-modes from the power spectrum when doing calculation according to Equation (\ref{aveDelta}). 

\begin{figure}	
        \begin{center}
        \includegraphics[scale=0.45]{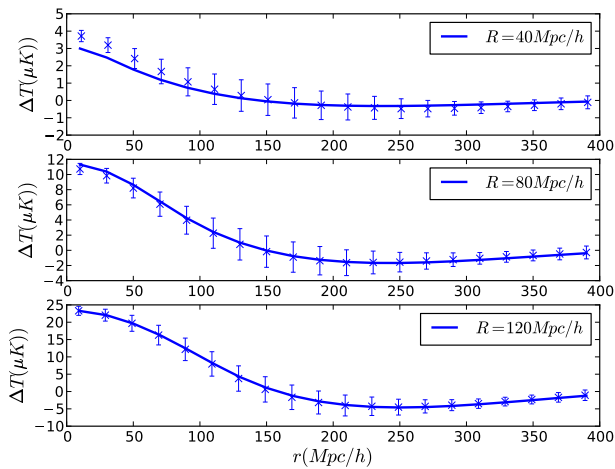}
        \end{center}
        \caption{The same as Figure \ref{iswprofHV} with the exception that the ISW map was filtered, not the density. The error bars represent the uncertainty of the measurement. \label{remove}}
\end{figure}

This approach has to be modified slightly when the CMB is given and late time anisotropies are affected by every mode of the galaxy distribution. The previous exercise still proves that low $k$-modes are to be ignored if we want to study the imprints of superstructures statistically. This can be achieved by filtering out these modes from the CMB before performing such analysis. In addition one has to remove the same components from the theoretical ISW profile before fitting the data. This is shown in Figure \ref{remove}. We filtered the ISW map and we filtered the ISW profile, which is implicitly expressed as $f(r_{2D},R)$, on the right side of Equation (\ref{tprof}), so that only modes with $|k|>2\pi\frac{5}{L}$ remained. After these manipulations the measurement remained consistent with linear theory.

The success of these tests proves the validity of the Gaussian model of \cite{Papai2010}. We studied Euclidean ISW projections because or their simplicity but there is no reason to think that our findings are restricted to the Euclidean case. The only major difference is that on a sphere one deals with spherical harmonics and instead of low $k$-modes we speak of low $l$-modes analogously.

\section{ISW Map of SDSS Superstructures}
\label{Match}

Having calculated the full linear ISW profile of superstructures in the dark matter density, we attempt to detect their presence on the CMB. With the locations of 50 superclusters and 50 supervoids found in the SDSS DR6 LRG sample by \cite{Granett2008} we created ISW maps and used the matched filter technique to measure the amplitude of the ISW signal. Our ISW maps are statistical in the sense that the individual ISW profiles were predicted by using statistical properties of Gaussian, spherically symmetric density fluctuations rather than the particular (poorly observed) realization of the density field.

\subsection{Superclusters and Supervoids in the SDSS LRG sample on $100\mpc$ scales}

Large regions with significantly high overdensities or low underdensities are called here superclusters or supervoids. In a given a dataset these regions can be defined in many different ways. Void finder algorithms were compared in \cite{Colberg2008}. Each implementation is based on a certain notion of a void. Voids can have spherical or irregular shapes, finders can be parameter free or there can be a hard coded density threshold or radius to describe what a void is like. It is important to emphasize that despite the similarities, any analysis using the output of a void finder will be sensitive to its particulars. This renders our study qualitative in nature. 

Our work is based on supervoids and superclusters found in a subsample of the SDSS DR6 LRG catalog by ZOBOV and VOBOZ, parameter free void and cluster finder algorithms \citep{Neyrinck2008,Neyrinck2005}. ZOBOV and VOBOZ calculate the Voronoi tessellation of the data in order to approximate the density field with a simple function, a funcion that is constant inside polyhedra. Voids are found by using the concept of drainage basins from Geography. Clusters are identified analogously.  

A table with the properties of 50 supervoids and 50 superclusters can be found in \cite{supplement}. Their sizes and densities cannot be quantified simply due to their shapes and survey boundary effects. The bias, redshift error and the shot noise are also sources of uncertainty for both size and density. This is why we used only the directions of the centers of these superstructures in creating ISW maps in Section \ref{MFT}. The error in the angular coordinates is relatively small.    

\subsection{CMB and ISW Maps}
\label{CaIM}

We used the Internal Linear Combination (ILC) Map from the WMAP 7-year release \citep{Gold2010} with the KQ75y7 galactic foreground and point source mask. We employed Healpix pixelization at resolution $n_{side} = 64$ \citep{Gorski2005}. In general, we strove to carry out the measurement with up to date maps and masks while remaining consistent with \cite{Granett2009}. 

Section \ref{PinS} concluded that a Gaussian model gave a good approximation to the expected ISW signal from a large spherical region in the dark matter field with given density contrast. It also became apparent that in an analysis of statistical ISW maps low $l$-modes were to be filtered out due to the large cosmic variance of the potential affecting these modes. This also lowers the cosmic variance. The lowest mode to be taken into account is determined by the sky coverage of the ISW map. Experimenting with the HV Simulation gave us $k_{min}\approx 2\pi\frac{5}{L}$ for an Euclidean ISW map, where $L$ is the linear size of the map. This translates into $l_{min}(l_{lmin}+1)\approx \frac{4\pi}{A}$, where $A$ is the angular size of ISW map. The SDSS DR6 imaging area about the north galactic pole covers roughly 8000 square degrees which gives us $l_{min} = 12$.

Due to the large uncertainty in their size and density we assume that each supervoid is a realization of the same statistical entity when we creating ISW maps. We consider superclusters similar in nature but with an opposite density contrast. 

\begin{figure}	
        \begin{center}
        \includegraphics[scale=0.45]{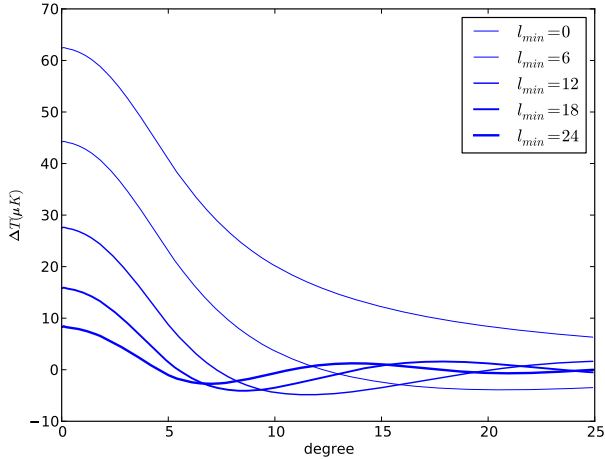}
        \end{center}
        \caption{The expected ISW imprint of a spherical region with radius of $120\mpc$ and an average $\delta$ of 1 at redshift 0.52. Each curve shows the same profile without modes under their respective $l_{min}$. \label{sphiswprof}}
\end{figure}

An ISW map is made simply by placing the expected ISW profiles of superstructures to the locations given in \cite{supplement}. We set $\delta_{in}$ to $1$ for clusters and $-1$ for voids in formulae of Section \ref{PinS}. For every radius there is a corresponding ISW map. We calculated the $a_{lm}$ coefficients of the spherical expansion of both the ISW maps and the CMB map, and zeroed the modes below $l_{min}=12$ before transforming the $a_{lm}$'s back. These manipulations were carried out with routines from the Healpix package \citep{Gorski2005}. Figure \ref{sphiswprof} shows how the ISW profile changes when modes are removed.   

\subsection{The Matched Filter Technique}
\label{MFT}

We continue with measuring the amplitude of the ISW maps created in Section \ref{CaIM} in the WMAP ILC map. To do this we used the same matched filter technique as \cite{Granett2009}. A likelihood function is defined as:
\begin{eqnarray}
L(\lambda) = -\frac{1}{2}\big( T_{cmb}-\lambda T_{ISW} \big)C^{-1}\big( T_{cmb}-\lambda T_{ISW} \big), \label{likelihood}
\end{eqnarray}   
where $T_{cmb}$ is a foreground cleaned CMB map, in this case the WMAP ILC map, and $T_{ISW}$ refers to the ISW maps. The pixel covariance matrix, $C$, is computed from the best fitting $\Lambda$CDM power spectrum. This likelihood is based on the assumption that the $T_{ISW}$ and the primary CMB are the realizations of two very different Gaussian random fields and the latter is given by the minimum of Equation (\ref{likelihood}):
\begin{eqnarray}
& \lambda^* = \frac{T_{ISW}C^{-1}T_{cmb}}{T_{ISW}C^{-1}T_{ISW}}, & \label{lambda} \\
& T_{prim} = T_{cmb} - \lambda^* T_{ISW}. &
\end{eqnarray}  
As explained in Section \ref{CaIM}, we filtered out modes up to $l_{min} = 12$ from both $T_{ISW}$ and $T_{cmb}$, and before building the covariance matrix from $C_{l}$'s we set $C_0 = C_1 =...=C_{11} = 0$.

Since the exact inverse of the covariance matrix is unstable for partial sky maps, we regularized $C$ by calculating a pseudo-inverse omitting the noisiest modes.        
The likelihood function above defines an error on $\lambda^*$ as 
\begin{eqnarray}
& \Delta \lambda^2 = \frac{1}{T_{ISW}C^{-1}T_{ISW}}\label{sigma}. &
\end{eqnarray}

\subsection{Results}
\label{results}

\begin{figure*}	
        \begin{center}
         \includegraphics[scale=0.4]{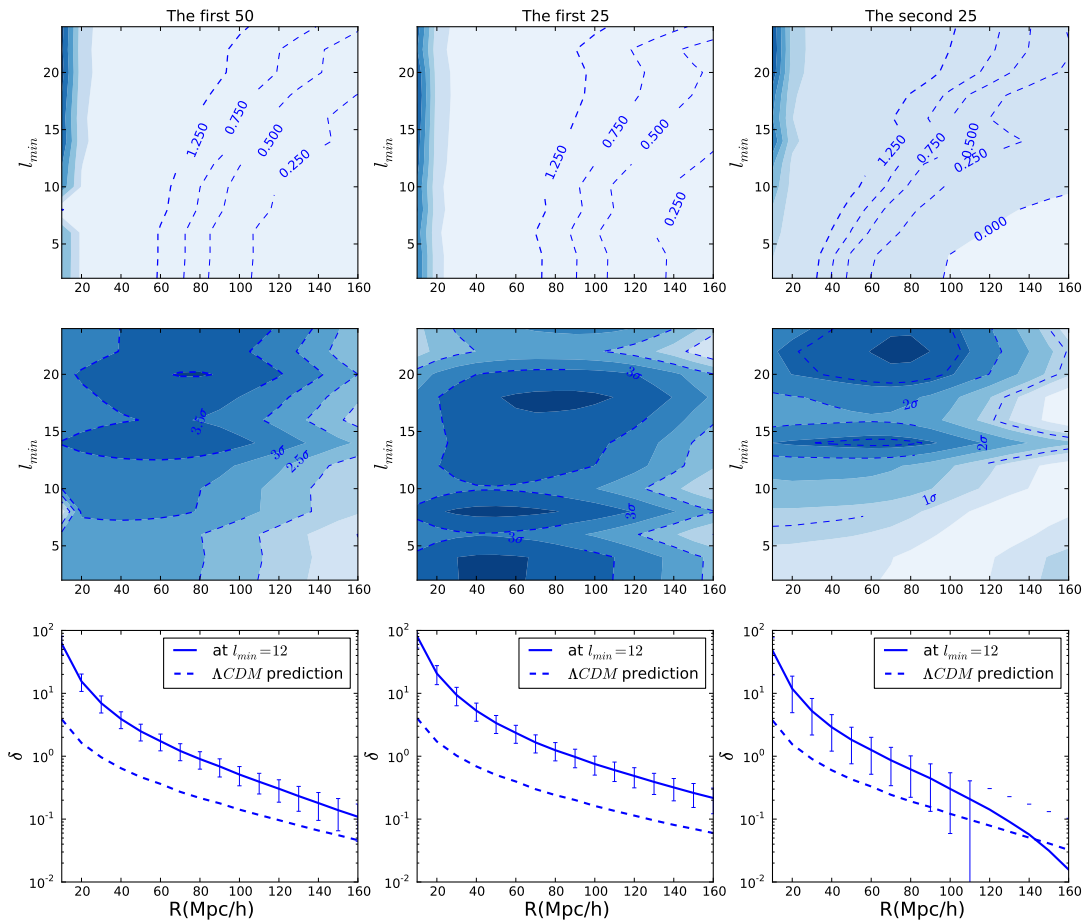}
        \end{center}
        \caption{The top and the mid row contain contour plots of the best fitting amplitude of the ISW maps, $\lambda$, and their relative uncertainty, $\sigma$, (see text for their definition) as functions of $l_{min}$ and $R$. The bottom row  shows cross sections of the contour plots from the top row at $l_{min} = 12$, as well as the prediction for the average density contrast of superstructures of a certain size. In the left column the fitting is for an ISW map consisting of the 50 most significant supervoids and 50 superclusters. The middle column shows the same for best 25-25 while the right for the remaining 50 superstructures not included in the middle column. \label{significance}}
\end{figure*}

We calculated $\lambda^*$ according to Equation (\ref{lambda}). Since we have ISW maps for different void/cluster radii, the result is a function, $\lambda(R)$. This is presented in Figure \ref{significance} in the bottom row on the left panel. The error bars are given by Equation (\ref{sigma}). From now on every time we refer to a $\lambda$ we mean the value at the minimum of the likelihood function, so we drop the superscript *. If we marginalize $\lambda(R)/\Delta \lambda(R)$ over $R$ we get a $3.24$-$\sigma$ detection. Similarly, the marginalized value of $R$ is $55 \pm 28 \mpc$. 

$\lambda(R)$ can be interpreted as the average density contrast in one hundred superstructures inside radius $R$, where the supervoids are taken into account with a minus sign: 
\begin{eqnarray}
& \lambda(R) \approx \frac{1}{2}\big( \delta_{in}^{50c}(R) - \delta_{in}^{50v}(R)\big), &
\end{eqnarray}
where superscripts $50c$ and $50v$ refer to the average of 50 superclusters and 50 supervoids. This interpretation originates from the linear relationship between the ISW signal and the density fluctuations, and our choice of $\delta_{in} = \pm 1$ when building ISW maps in Section \ref{CaIM}. In addition to the data curves, in Figure \ref{significance} we plotted the prediction of linear theory for the average of the 50 largest superclusters. We assumed a Gaussian Probability Distribution Function (PDF) which is reasonable in light of the findings of \cite{Papai2010}. The variance of the PDF was calculated by CAMB \citep{lewis} using the latest WMAP cosmological parameters \citep{Jarosik2010} and it was scaled with the growth function to the median redshift of the superstructures $(z=0.52)$. The number of independent clusters was estimated as $V_{survey}/V_{cluster}$ where $V_{survey}$ is the volume of the LRG sample used by \cite{Granett2009}. $\lambda$ is consistently several times larger than the predicted $\delta_{in}$. As a consistency test we split the voids and clusters into two groups. Group 1 consists of the first 25 voids and 25 clusters with the highest significance in the supplementary tables of \cite{supplement}. The other 25 voids and 25 clusters make up group 2. We created two ISW maps, one from each of the groups. After repeating the whole procedure for these, we got the center and the right panel in the bottom row of Figure \ref{significance}. As expected, superstructures associated with lower significance produce lower signal on the CMB and allow larger deviation from the average $\lambda$. This supports the hypothesis that the signal is due to the ISW effect.

We also checked the sensitivity of the results to the choice of $l_{min}$, the lowest spherical harmonics included in the analysis. The result for different choices of $l_{min}$ is shown in the mid and top rows of Figure \ref{significance}. The top row shows $\lambda(R,l_{min})$, while the mid row shows $\sigma(R,l_{min}) = \lambda(R,l_{min})/\Delta \lambda(R,l_{min})$ on contour plots. The result seems robust for wide range of $l_{min}$'s. 

\begin{figure}	
        \begin{center}
        \includegraphics[scale=0.45]{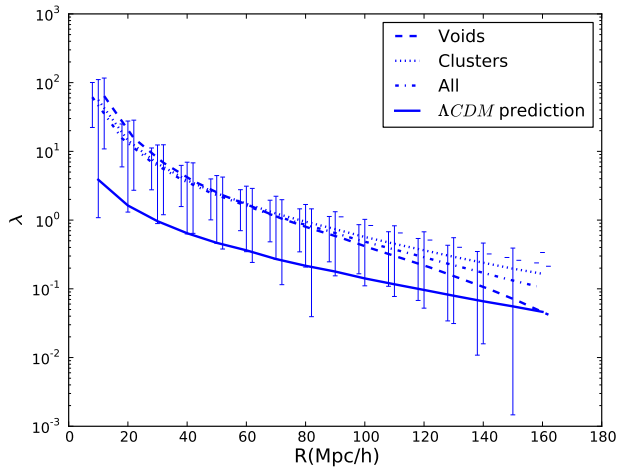}
        \end{center}
        \caption{After fitting simple ISW maps made from the expected imprint of single superclusters or supervoids we were left with 50-50 best amplitudes, $\lambda$'s. Their averages with the $2\sigma$ uncertainty are plotted with the amplitude of a full ISW map consisting of all 100 imprints. These measurements were done with $l_{min} = 12$. The theoretical prediction (continuous line) for the density contrast is plotted along. \label{single}}
\end{figure}

We performed another consistency check to acquire a better understanding of the uncertainty of $\lambda$ and its relationship with $\delta_{in}$. We created ISW maps from single superstructures. We used the matched filter technique to determine a $\lambda$ and its error as before. The average $\lambda$ for voids and for clusters separately are plotted in Figure \ref{single}. In this case we defined the uncertainty of $\lambda$ as the standard deviation of its mean. We also plotted the previous results for the previous combined ISW map. All of this is for $l_{min} = 12$. The three measured curves are consistent with each other, the error bars show the two $\sigma$ deviation. The single void and cluster fit allow wider range of $\lambda$ than the combined fit.

\section{Discussion}
\label{discussion}

This paper is the continuation of the work started in \cite{Papai2010}. Our goal is to estimate the ISW imprint of large spherical dark matter overdensities and underdensities. In Section \ref{PinS}, first, we tested a simple Gaussian model for density profiles in the HVS. We created an ISW map by ray tracing through the simulation and computing the linear ISW effect. We neglected the nonlinear part of the ISW effect, as it is small in comparison at $z=0$ \citep{Cai2010}. Another simplification was that our ISW map was Euclidean. For the purpose of testing this has no relevance. 

As it can be seen from Figure \ref{HVsig2}, the density profiles followed linear theory given by Equation (\ref{aveDelta}) within $1$-$\sigma$ uncertainty up to $400\mpc$, the largest scale in our study. This is not surprising, since the density profile is equivalent of the two-point function and on large scales high-order clustering is not important. When we averaged the density profiles we only used locations where the average density contrast in a certain radius was in the $2$-$\sigma$ (positive or negative) tail  of the PDF. This demonstrates that not even the extreme cases are affected by nonlinearities significantly. 

We demonstrated that cosmic variance was a much more important factor for the ISW effect than for the density. Since cosmic variance is large for low $k$-modes and the potential is non-vanishing as $k\rightarrow 0$, these modes have relatively large effect on the ISW profile. Surely large variance modes should be removed from the CMB and the ISW maps if the ISW maps are based on the statistics of density fluctuations and not on the observed density. First we measured the ISW profiles after filtering out the low $k$-modes from the HVS density and consequently the ISW map. Filtering the two-point function in the same way gave a good match to the data. (See Figure \ref{iswprofHV}.) 

If the ISW map is given, as in the CMB, another route has to be taken. We filtered the complete ISW map and compared it to the theoretical profile after removing the same modes. The result is shown in Figure \ref{remove}    

In Section \ref{Match} we used matched filter technique to detect the ISW signal of superstructures in the CMB. We closely followed the procedure described in \cite{Granett2009}. The major difference is that our ISW map was built from theoretical assumptions about the shape of ISW profiles, while in \cite{Granett2009} the authors used an analytic curve which fitted the measured profiles the best. Because of this, the significance of our measurement can readily be interpreted. We estimated the marginalized significance of our measurement to be $\sigma = 3.24$. (See Subsection \ref{results} for details.) The interpretation of the best fitting amplitude of the ISW map is still not without difficulties. It is several times higher than the anticipated signal (Figures \ref{significance} and \ref{single}), although it appears to be $2$-$\sigma$ higher than $\Lambda$CDM predictions. Despite the difference in amplitude, the theoretical and the measured curves run parallel with each other, which supports that the signal is related to the ISW effect. We caution though that the uncertainty can still be underestimated, since it was derived from data alone. 

The ISW effect provides an explanation for another feature of observations. On average supervoids tend to have a hot ring around an initial cold dip and the opposite is true for superclusters \citep{Granett2009}. 
When we erased large scale fluctuations in the microwave background, the expected ISW profile changed sign as seen in Figure \ref{sphiswprof}. This means that in certain, not particularly unique realizations the average profiles of 50 supervoids or superstructures from a finite area of the sky can have rings in the ISW context, simply due to the cosmic variance of low $k$-modes.

Overall we can say that the shape of the measured signal follows the predicted ISW profile while its amplitude exceeds expectation. This is a good reason to investigate further by studying galaxy surveys other than the SDSS. A key to a quantitative study is the well-measured galaxy density. Since the model described in this paper is a model for density fluctuations around a particular location, which is not necessarily a maximum, it is enough to know the density at this location accurately.  

We thank Mark Neyrinck for sharing his knowledge on void and cluster finders. The authors were supported by NASA grants NNX10AD53G and NNG06GE71G, and the Pol\'anyi program of the Hungarian National Office for the Research and Technology (NKTH).

The Hubble Volume Simulations were carried out by the Virgo Supercomputing Consortium using computers based at the Computing Centre of the Max-Planck Society in Garching and at the Edinburgh parallel Computing Centre. The data are publicly available at http://www.mpa-garching.mpg.de/NumCos.

\bibliographystyle{apsrmp}
\bibliography{ms}

\begin{thebibliography}{32}
\expandafter\ifx\csname natexlab\endcsname\relax\def\natexlab#1{#1}\fi
\expandafter\ifx\csname bibnamefont\endcsname\relax
  \def\bibnamefont#1{#1}\fi
\expandafter\ifx\csname bibfnamefont\endcsname\relax
  \def\bibfnamefont#1{#1}\fi
\expandafter\ifx\csname citenamefont\endcsname\relax
  \def\citenamefont#1{#1}\fi
\expandafter\ifx\csname url\endcsname\relax
  \def\url#1{\texttt{#1}}\fi
\expandafter\ifx\csname urlprefix\endcsname\relax\def\urlprefix{URL }\fi
\providecommand{\bibinfo}[2]{#2}
\providecommand{\eprint}[2][]{\url{#2}}

\bibitem[{\citenamefont{{Afshordi}}
  \emph{et~al.}(2004)\citenamefont{{Afshordi}, {Loh}, and
  {Strauss}}}]{Afshordi2004}
\bibinfo{author}{\bibnamefont{{Afshordi}}, \bibfnamefont{N.}},
  \bibinfo{author}{\bibfnamefont{Y.}~\bibnamefont{{Loh}}}, and
  \bibinfo{author}{\bibfnamefont{M.~A.} \bibnamefont{{Strauss}}},
  \bibinfo{year}{2004}, \bibinfo{journal}{\prd}
  \textbf{\bibinfo{volume}{69}}(\bibinfo{number}{8}), \bibinfo{pages}{083524}.

\bibitem[{\citenamefont{{Afshordi} and {Tolley}}(2008)}]{Afshordi2008}
\bibinfo{author}{\bibnamefont{{Afshordi}}, \bibfnamefont{N.}}, and
  \bibinfo{author}{\bibfnamefont{A.~J.} \bibnamefont{{Tolley}}},
  \bibinfo{year}{2008}, \bibinfo{journal}{\prd}
  \textbf{\bibinfo{volume}{78}}(\bibinfo{number}{12}), \bibinfo{pages}{123507}.

\bibitem[{\citenamefont{{Bardeen}} \emph{et~al.}(1986)\citenamefont{{Bardeen},
  {Bond}, {Kaiser}, and {Szalay}}}]{BBKS1986}
\bibinfo{author}{\bibnamefont{{Bardeen}}, \bibfnamefont{J.~M.}},
  \bibinfo{author}{\bibfnamefont{J.~R.} \bibnamefont{{Bond}}},
  \bibinfo{author}{\bibfnamefont{N.}~\bibnamefont{{Kaiser}}}, and
  \bibinfo{author}{\bibfnamefont{A.~S.} \bibnamefont{{Szalay}}},
  \bibinfo{year}{1986}, \bibinfo{journal}{\apj} \textbf{\bibinfo{volume}{304}},
  \bibinfo{pages}{15}.

\bibitem[{\citenamefont{{Cai}} \emph{et~al.}(2010)\citenamefont{{Cai}, {Cole},
  {Jenkins}, and {Frenk}}}]{Cai2010}
\bibinfo{author}{\bibnamefont{{Cai}}, \bibfnamefont{Y.}},
  \bibinfo{author}{\bibfnamefont{S.}~\bibnamefont{{Cole}}},
  \bibinfo{author}{\bibfnamefont{A.}~\bibnamefont{{Jenkins}}}, and
  \bibinfo{author}{\bibfnamefont{C.~S.} \bibnamefont{{Frenk}}},
  \bibinfo{year}{2010}, \bibinfo{journal}{\mnras}
  \textbf{\bibinfo{volume}{407}}, \bibinfo{pages}{201}.

\bibitem[{\citenamefont{{Colberg}} \emph{et~al.}(2008)\citenamefont{{Colberg},
  {Pearce}, {Foster}, {Platen}, {Brunino}, {Neyrinck}, {Basilakos}, {Fairall},
  {Feldman}, {Gottl{\"o}ber}, {Hahn}, {Hoyle}} \emph{et~al.}}]{Colberg2008}
\bibinfo{author}{\bibnamefont{{Colberg}}, \bibfnamefont{J.~M.}},
  \bibinfo{author}{\bibfnamefont{F.}~\bibnamefont{{Pearce}}},
  \bibinfo{author}{\bibfnamefont{C.}~\bibnamefont{{Foster}}},
  \bibinfo{author}{\bibfnamefont{E.}~\bibnamefont{{Platen}}},
  \bibinfo{author}{\bibfnamefont{R.}~\bibnamefont{{Brunino}}},
  \bibinfo{author}{\bibfnamefont{M.}~\bibnamefont{{Neyrinck}}},
  \bibinfo{author}{\bibfnamefont{S.}~\bibnamefont{{Basilakos}}},
  \bibinfo{author}{\bibfnamefont{A.}~\bibnamefont{{Fairall}}},
  \bibinfo{author}{\bibfnamefont{H.}~\bibnamefont{{Feldman}}},
  \bibinfo{author}{\bibfnamefont{S.}~\bibnamefont{{Gottl{\"o}ber}}},
  \bibinfo{author}{\bibfnamefont{O.}~\bibnamefont{{Hahn}}},
  \bibinfo{author}{\bibfnamefont{F.}~\bibnamefont{{Hoyle}}}, \emph{et~al.},
  \bibinfo{year}{2008}, \bibinfo{journal}{\mnras}
  \textbf{\bibinfo{volume}{387}}, \bibinfo{pages}{933}.

\bibitem[{\citenamefont{{Colberg}} \emph{et~al.}(2000)\citenamefont{{Colberg},
  {White}, {Yoshida}, {MacFarland}, {Jenkins}, {Frenk}, {Pearce}, {Evrard},
  {Couchman}, {Efstathiou}, {Peacock}, {Thomas}} \emph{et~al.}}]{Colberg2000}
\bibinfo{author}{\bibnamefont{{Colberg}}, \bibfnamefont{J.~M.}},
  \bibinfo{author}{\bibfnamefont{S.~D.~M.} \bibnamefont{{White}}},
  \bibinfo{author}{\bibfnamefont{N.}~\bibnamefont{{Yoshida}}},
  \bibinfo{author}{\bibfnamefont{T.~J.} \bibnamefont{{MacFarland}}},
  \bibinfo{author}{\bibfnamefont{A.}~\bibnamefont{{Jenkins}}},
  \bibinfo{author}{\bibfnamefont{C.~S.} \bibnamefont{{Frenk}}},
  \bibinfo{author}{\bibfnamefont{F.~R.} \bibnamefont{{Pearce}}},
  \bibinfo{author}{\bibfnamefont{A.~E.} \bibnamefont{{Evrard}}},
  \bibinfo{author}{\bibfnamefont{H.~M.~P.} \bibnamefont{{Couchman}}},
  \bibinfo{author}{\bibfnamefont{G.}~\bibnamefont{{Efstathiou}}},
  \bibinfo{author}{\bibfnamefont{J.~A.} \bibnamefont{{Peacock}}},
  \bibinfo{author}{\bibfnamefont{P.~A.} \bibnamefont{{Thomas}}}, \emph{et~al.},
  \bibinfo{year}{2000}, \bibinfo{journal}{\mnras}
  \textbf{\bibinfo{volume}{319}}, \bibinfo{pages}{209}.

\bibitem[{\citenamefont{{Crocce}} \emph{et~al.}(2006)\citenamefont{{Crocce},
  {Pueblas}, and {Scoccimarro}}}]{Crocce2006}
\bibinfo{author}{\bibnamefont{{Crocce}}, \bibfnamefont{M.}},
  \bibinfo{author}{\bibfnamefont{S.}~\bibnamefont{{Pueblas}}}, and
  \bibinfo{author}{\bibfnamefont{R.}~\bibnamefont{{Scoccimarro}}},
  \bibinfo{year}{2006}, \bibinfo{journal}{\mnras}
  \textbf{\bibinfo{volume}{373}}, \bibinfo{pages}{369}.

\bibitem[{\citenamefont{{Cruz}} \emph{et~al.}(2005)\citenamefont{{Cruz},
  {Mart{\'{\i}}nez-Gonz{\'a}lez}, {Vielva}, and {Cay{\'o}n}}}]{Cruz2005}
\bibinfo{author}{\bibnamefont{{Cruz}}, \bibfnamefont{M.}},
  \bibinfo{author}{\bibfnamefont{E.}~\bibnamefont{{Mart{\'{\i}}nez-Gonz{\'a}lez}}},
  \bibinfo{author}{\bibfnamefont{P.}~\bibnamefont{{Vielva}}}, and
  \bibinfo{author}{\bibfnamefont{L.}~\bibnamefont{{Cay{\'o}n}}},
  \bibinfo{year}{2005}, \bibinfo{journal}{\mnras}
  \textbf{\bibinfo{volume}{356}}, \bibinfo{pages}{29}.

\bibitem[{\citenamefont{{Cruz}} \emph{et~al.}(2007)\citenamefont{{Cruz},
  {Turok}, {Vielva}, {Mart{\'{\i}}nez-Gonz{\'a}lez}, and {Hobson}}}]{Cruz2007}
\bibinfo{author}{\bibnamefont{{Cruz}}, \bibfnamefont{M.}},
  \bibinfo{author}{\bibfnamefont{N.}~\bibnamefont{{Turok}}},
  \bibinfo{author}{\bibfnamefont{P.}~\bibnamefont{{Vielva}}},
  \bibinfo{author}{\bibfnamefont{E.}~\bibnamefont{{Mart{\'{\i}}nez-Gonz{\'a}lez}}},
  and \bibinfo{author}{\bibfnamefont{M.}~\bibnamefont{{Hobson}}},
  \bibinfo{year}{2007}, \bibinfo{journal}{Science}
  \textbf{\bibinfo{volume}{318}}, \bibinfo{pages}{1612}.

\bibitem[{\citenamefont{{Dekel}}(1981)}]{Dekel1981}
\bibinfo{author}{\bibnamefont{{Dekel}}, \bibfnamefont{A.}},
  \bibinfo{year}{1981}, \bibinfo{journal}{\aap} \textbf{\bibinfo{volume}{101}},
  \bibinfo{pages}{79}.

\bibitem[{\citenamefont{{Giannantonio}}
  \emph{et~al.}(2008{\natexlab{a}})\citenamefont{{Giannantonio}, {Scranton},
  {Crittenden}, {Nichol}, {Boughn}, {Myers}, and
  {Richards}}}]{Giannantonio2008}
\bibinfo{author}{\bibnamefont{{Giannantonio}}, \bibfnamefont{T.}},
  \bibinfo{author}{\bibfnamefont{R.}~\bibnamefont{{Scranton}}},
  \bibinfo{author}{\bibfnamefont{R.~G.} \bibnamefont{{Crittenden}}},
  \bibinfo{author}{\bibfnamefont{R.~C.} \bibnamefont{{Nichol}}},
  \bibinfo{author}{\bibfnamefont{S.~P.} \bibnamefont{{Boughn}}},
  \bibinfo{author}{\bibfnamefont{A.~D.} \bibnamefont{{Myers}}}, and
  \bibinfo{author}{\bibfnamefont{G.~T.} \bibnamefont{{Richards}}},
  \bibinfo{year}{2008}{\natexlab{a}}, \bibinfo{journal}{\prd}
  \textbf{\bibinfo{volume}{77}}(\bibinfo{number}{12}), \bibinfo{pages}{123520}.

\bibitem[{\citenamefont{{Giannantonio}}
  \emph{et~al.}(2008{\natexlab{b}})\citenamefont{{Giannantonio}, {Song}, and
  {Koyama}}}]{Giannantonio2008b}
\bibinfo{author}{\bibnamefont{{Giannantonio}}, \bibfnamefont{T.}},
  \bibinfo{author}{\bibfnamefont{Y.}~\bibnamefont{{Song}}}, and
  \bibinfo{author}{\bibfnamefont{K.}~\bibnamefont{{Koyama}}},
  \bibinfo{year}{2008}{\natexlab{b}}, \bibinfo{journal}{\prd}
  \textbf{\bibinfo{volume}{78}}(\bibinfo{number}{4}), \bibinfo{pages}{044017}.

\bibitem[{\citenamefont{{Gold}} \emph{et~al.}(2010)\citenamefont{{Gold},
  {Odegard}, {Weiland}, {Hill}, {Kogut}, {Bennett}, {Hinshaw}, {Chen},
  {Dunkley}, {Halpern}, {Jarosik}, {Komatsu}} \emph{et~al.}}]{Gold2010}
\bibinfo{author}{\bibnamefont{{Gold}}, \bibfnamefont{B.}},
  \bibinfo{author}{\bibfnamefont{N.}~\bibnamefont{{Odegard}}},
  \bibinfo{author}{\bibfnamefont{J.~L.} \bibnamefont{{Weiland}}},
  \bibinfo{author}{\bibfnamefont{R.~S.} \bibnamefont{{Hill}}},
  \bibinfo{author}{\bibfnamefont{A.}~\bibnamefont{{Kogut}}},
  \bibinfo{author}{\bibfnamefont{C.~L.} \bibnamefont{{Bennett}}},
  \bibinfo{author}{\bibfnamefont{G.}~\bibnamefont{{Hinshaw}}},
  \bibinfo{author}{\bibfnamefont{X.}~\bibnamefont{{Chen}}},
  \bibinfo{author}{\bibfnamefont{J.}~\bibnamefont{{Dunkley}}},
  \bibinfo{author}{\bibfnamefont{M.}~\bibnamefont{{Halpern}}},
  \bibinfo{author}{\bibfnamefont{N.}~\bibnamefont{{Jarosik}}},
  \bibinfo{author}{\bibfnamefont{E.}~\bibnamefont{{Komatsu}}}, \emph{et~al.},
  \bibinfo{year}{2010}, \bibinfo{journal}{ArXiv e-prints} \eprint{1001.4555}.

\bibitem[{\citenamefont{{G{\'o}rski}}
  \emph{et~al.}(2005)\citenamefont{{G{\'o}rski}, {Hivon}, {Banday}, {Wandelt},
  {Hansen}, {Reinecke}, and {Bartelmann}}}]{Gorski2005}
\bibinfo{author}{\bibnamefont{{G{\'o}rski}}, \bibfnamefont{K.~M.}},
  \bibinfo{author}{\bibfnamefont{E.}~\bibnamefont{{Hivon}}},
  \bibinfo{author}{\bibfnamefont{A.~J.} \bibnamefont{{Banday}}},
  \bibinfo{author}{\bibfnamefont{B.~D.} \bibnamefont{{Wandelt}}},
  \bibinfo{author}{\bibfnamefont{F.~K.} \bibnamefont{{Hansen}}},
  \bibinfo{author}{\bibfnamefont{M.}~\bibnamefont{{Reinecke}}}, and
  \bibinfo{author}{\bibfnamefont{M.}~\bibnamefont{{Bartelmann}}},
  \bibinfo{year}{2005}, \bibinfo{journal}{\apj} \textbf{\bibinfo{volume}{622}},
  \bibinfo{pages}{759}.

\bibitem[{\citenamefont{{Granett}}
  \emph{et~al.}(2008{\natexlab{a}})\citenamefont{{Granett}, {Neyrinck}, and
  {Szapudi}}}]{Granett2008}
\bibinfo{author}{\bibnamefont{{Granett}}, \bibfnamefont{B.~R.}},
  \bibinfo{author}{\bibfnamefont{M.~C.} \bibnamefont{{Neyrinck}}}, and
  \bibinfo{author}{\bibfnamefont{I.}~\bibnamefont{{Szapudi}}},
  \bibinfo{year}{2008}{\natexlab{a}}, \bibinfo{journal}{\apjl}
  \textbf{\bibinfo{volume}{683}}, \bibinfo{pages}{L99}.

\bibitem[{\citenamefont{{Granett}}
  \emph{et~al.}(2008{\natexlab{b}})\citenamefont{{Granett}, {Neyrinck}, and
  {Szapudi}}}]{supplement}
\bibinfo{author}{\bibnamefont{{Granett}}, \bibfnamefont{B.~R.}},
  \bibinfo{author}{\bibfnamefont{M.~C.} \bibnamefont{{Neyrinck}}}, and
  \bibinfo{author}{\bibfnamefont{I.}~\bibnamefont{{Szapudi}}},
  \bibinfo{year}{2008}{\natexlab{b}}, \bibinfo{journal}{ArXiv e-prints}
  \eprint{0805.2974}.

\bibitem[{\citenamefont{{Granett}} \emph{et~al.}(2009)\citenamefont{{Granett},
  {Neyrinck}, and {Szapudi}}}]{Granett2009}
\bibinfo{author}{\bibnamefont{{Granett}}, \bibfnamefont{B.~R.}},
  \bibinfo{author}{\bibfnamefont{M.~C.} \bibnamefont{{Neyrinck}}}, and
  \bibinfo{author}{\bibfnamefont{I.}~\bibnamefont{{Szapudi}}},
  \bibinfo{year}{2009}, \bibinfo{journal}{\apj} \textbf{\bibinfo{volume}{701}},
  \bibinfo{pages}{414}.

\bibitem[{\citenamefont{{Gurzadyan} and {Penrose}}(2010)}]{bs}
\bibinfo{author}{\bibnamefont{{Gurzadyan}}, \bibfnamefont{V.~G.}}, and
  \bibinfo{author}{\bibfnamefont{R.}~\bibnamefont{{Penrose}}},
  \bibinfo{year}{2010}, \bibinfo{journal}{ArXiv e-prints} \eprint{1011.3706}.

\bibitem[{\citenamefont{{Ho}} \emph{et~al.}(2008)\citenamefont{{Ho}, {Hirata},
  {Padmanabhan}, {Seljak}, and {Bahcall}}}]{Ho2008}
\bibinfo{author}{\bibnamefont{{Ho}}, \bibfnamefont{S.}},
  \bibinfo{author}{\bibfnamefont{C.}~\bibnamefont{{Hirata}}},
  \bibinfo{author}{\bibfnamefont{N.}~\bibnamefont{{Padmanabhan}}},
  \bibinfo{author}{\bibfnamefont{U.}~\bibnamefont{{Seljak}}}, and
  \bibinfo{author}{\bibfnamefont{N.}~\bibnamefont{{Bahcall}}},
  \bibinfo{year}{2008}, \bibinfo{journal}{\prd}
  \textbf{\bibinfo{volume}{78}}(\bibinfo{number}{4}), \bibinfo{pages}{043519}.

\bibitem[{\citenamefont{{Inoue}} \emph{et~al.}(2010)\citenamefont{{Inoue},
  {Sakai}, and {Tomita}}}]{Inoue2010}
\bibinfo{author}{\bibnamefont{{Inoue}}, \bibfnamefont{K.~T.}},
  \bibinfo{author}{\bibfnamefont{N.}~\bibnamefont{{Sakai}}}, and
  \bibinfo{author}{\bibfnamefont{K.}~\bibnamefont{{Tomita}}},
  \bibinfo{year}{2010}, \bibinfo{journal}{\apj} \textbf{\bibinfo{volume}{724}},
  \bibinfo{pages}{12}.

\bibitem[{\citenamefont{{Jarosik}} \emph{et~al.}(2010)\citenamefont{{Jarosik},
  {Bennett}, {Dunkley}, {Gold}, {Greason}, {Halpern}, {Hill}, {Hinshaw},
  {Kogut}, {Komatsu}, {Larson}, {Limon}} \emph{et~al.}}]{Jarosik2010}
\bibinfo{author}{\bibnamefont{{Jarosik}}, \bibfnamefont{N.}},
  \bibinfo{author}{\bibfnamefont{C.~L.} \bibnamefont{{Bennett}}},
  \bibinfo{author}{\bibfnamefont{J.}~\bibnamefont{{Dunkley}}},
  \bibinfo{author}{\bibfnamefont{B.}~\bibnamefont{{Gold}}},
  \bibinfo{author}{\bibfnamefont{M.~R.} \bibnamefont{{Greason}}},
  \bibinfo{author}{\bibfnamefont{M.}~\bibnamefont{{Halpern}}},
  \bibinfo{author}{\bibfnamefont{R.~S.} \bibnamefont{{Hill}}},
  \bibinfo{author}{\bibfnamefont{G.}~\bibnamefont{{Hinshaw}}},
  \bibinfo{author}{\bibfnamefont{A.}~\bibnamefont{{Kogut}}},
  \bibinfo{author}{\bibfnamefont{E.}~\bibnamefont{{Komatsu}}},
  \bibinfo{author}{\bibfnamefont{D.}~\bibnamefont{{Larson}}},
  \bibinfo{author}{\bibfnamefont{M.}~\bibnamefont{{Limon}}}, \emph{et~al.},
  \bibinfo{year}{2010}, \bibinfo{journal}{ArXiv e-prints} \eprint{1001.4744}.

\bibitem[{\citenamefont{{Lewis}} \emph{et~al.}(2000)\citenamefont{{Lewis},
  {Challinor}, and {Lasenby}}}]{lewis}
\bibinfo{author}{\bibnamefont{{Lewis}}, \bibfnamefont{A.}},
  \bibinfo{author}{\bibfnamefont{A.}~\bibnamefont{{Challinor}}}, and
  \bibinfo{author}{\bibfnamefont{A.}~\bibnamefont{{Lasenby}}},
  \bibinfo{year}{2000}, \bibinfo{journal}{\apj} \textbf{\bibinfo{volume}{538}},
  \bibinfo{pages}{473}.

\bibitem[{\citenamefont{{Neyrinck}}(2008)}]{Neyrinck2008}
\bibinfo{author}{\bibnamefont{{Neyrinck}}, \bibfnamefont{M.~C.}},
  \bibinfo{year}{2008}, \bibinfo{journal}{\mnras}
  \textbf{\bibinfo{volume}{386}}, \bibinfo{pages}{2101}.

\bibitem[{\citenamefont{{Neyrinck}}
  \emph{et~al.}(2005)\citenamefont{{Neyrinck}, {Gnedin}, and
  {Hamilton}}}]{Neyrinck2005}
\bibinfo{author}{\bibnamefont{{Neyrinck}}, \bibfnamefont{M.~C.}},
  \bibinfo{author}{\bibfnamefont{N.~Y.} \bibnamefont{{Gnedin}}}, and
  \bibinfo{author}{\bibfnamefont{A.~J.~S.} \bibnamefont{{Hamilton}}},
  \bibinfo{year}{2005}, \bibinfo{journal}{\mnras}
  \textbf{\bibinfo{volume}{356}}, \bibinfo{pages}{1222}.

\bibitem[{\citenamefont{{Padmanabhan}}
  \emph{et~al.}(2005)\citenamefont{{Padmanabhan}, {Hirata}, {Seljak},
  {Schlegel}, {Brinkmann}, and {Schneider}}}]{Padmanabhan2005}
\bibinfo{author}{\bibnamefont{{Padmanabhan}}, \bibfnamefont{N.}},
  \bibinfo{author}{\bibfnamefont{C.~M.} \bibnamefont{{Hirata}}},
  \bibinfo{author}{\bibfnamefont{U.}~\bibnamefont{{Seljak}}},
  \bibinfo{author}{\bibfnamefont{D.~J.} \bibnamefont{{Schlegel}}},
  \bibinfo{author}{\bibfnamefont{J.}~\bibnamefont{{Brinkmann}}}, and
  \bibinfo{author}{\bibfnamefont{D.~P.} \bibnamefont{{Schneider}}},
  \bibinfo{year}{2005}, \bibinfo{journal}{\prd}
  \textbf{\bibinfo{volume}{72}}(\bibinfo{number}{4}), \bibinfo{pages}{043525}.

\bibitem[{\citenamefont{{P{\'a}pai} and {Szapudi}}(2010)}]{Papai2010}
\bibinfo{author}{\bibnamefont{{P{\'a}pai}}, \bibfnamefont{P.}}, and
  \bibinfo{author}{\bibfnamefont{I.}~\bibnamefont{{Szapudi}}},
  \bibinfo{year}{2010}, \bibinfo{journal}{\apj} \textbf{\bibinfo{volume}{725}},
  \bibinfo{pages}{2078}.

\bibitem[{\citenamefont{{Raccanelli}}
  \emph{et~al.}(2008)\citenamefont{{Raccanelli}, {Bonaldi}, {Negrello},
  {Matarrese}, {Tormen}, and {de Zotti}}}]{Raccanelli2008}
\bibinfo{author}{\bibnamefont{{Raccanelli}}, \bibfnamefont{A.}},
  \bibinfo{author}{\bibfnamefont{A.}~\bibnamefont{{Bonaldi}}},
  \bibinfo{author}{\bibfnamefont{M.}~\bibnamefont{{Negrello}}},
  \bibinfo{author}{\bibfnamefont{S.}~\bibnamefont{{Matarrese}}},
  \bibinfo{author}{\bibfnamefont{G.}~\bibnamefont{{Tormen}}}, and
  \bibinfo{author}{\bibfnamefont{G.}~\bibnamefont{{de Zotti}}},
  \bibinfo{year}{2008}, \bibinfo{journal}{\mnras}
  \textbf{\bibinfo{volume}{386}}, \bibinfo{pages}{2161}.

\bibitem[{\citenamefont{{Rudnick}} \emph{et~al.}(2007)\citenamefont{{Rudnick},
  {Brown}, and {Williams}}}]{Rudnick2007}
\bibinfo{author}{\bibnamefont{{Rudnick}}, \bibfnamefont{L.}},
  \bibinfo{author}{\bibfnamefont{S.}~\bibnamefont{{Brown}}}, and
  \bibinfo{author}{\bibfnamefont{L.~R.} \bibnamefont{{Williams}}},
  \bibinfo{year}{2007}, \bibinfo{journal}{\apj} \textbf{\bibinfo{volume}{671}},
  \bibinfo{pages}{40}.

\bibitem[{\citenamefont{{Sachs} and {Wolfe}}(1967)}]{SW1967}
\bibinfo{author}{\bibnamefont{{Sachs}}, \bibfnamefont{R.~K.}}, and
  \bibinfo{author}{\bibfnamefont{A.~M.} \bibnamefont{{Wolfe}}},
  \bibinfo{year}{1967}, \bibinfo{journal}{\apj} \textbf{\bibinfo{volume}{147}},
  \bibinfo{pages}{73}.

\bibitem[{\citenamefont{{Sawangwit}}
  \emph{et~al.}(2010)\citenamefont{{Sawangwit}, {Shanks}, {Cannon}, {Croom},
  {Ross}, and {Wake}}}]{Sawangwit2010}
\bibinfo{author}{\bibnamefont{{Sawangwit}}, \bibfnamefont{U.}},
  \bibinfo{author}{\bibfnamefont{T.}~\bibnamefont{{Shanks}}},
  \bibinfo{author}{\bibfnamefont{R.~D.} \bibnamefont{{Cannon}}},
  \bibinfo{author}{\bibfnamefont{S.~M.} \bibnamefont{{Croom}}},
  \bibinfo{author}{\bibfnamefont{N.~P.} \bibnamefont{{Ross}}}, and
  \bibinfo{author}{\bibfnamefont{D.~A.} \bibnamefont{{Wake}}},
  \bibinfo{year}{2010}, \bibinfo{journal}{\mnras}
  \textbf{\bibinfo{volume}{402}}, \bibinfo{pages}{2228}.

\bibitem[{\citenamefont{{Scranton}}
  \emph{et~al.}(2003)\citenamefont{{Scranton}, {Connolly}, {Nichol},
  {Stebbins}, {Szapudi}, {Eisenstein}, {Afshordi}, {Budavari}, {Csabai},
  {Frieman}, {Gunn}, {Johnson}} \emph{et~al.}}]{Scranton2003}
\bibinfo{author}{\bibnamefont{{Scranton}}, \bibfnamefont{R.}},
  \bibinfo{author}{\bibfnamefont{A.~J.} \bibnamefont{{Connolly}}},
  \bibinfo{author}{\bibfnamefont{R.~C.} \bibnamefont{{Nichol}}},
  \bibinfo{author}{\bibfnamefont{A.}~\bibnamefont{{Stebbins}}},
  \bibinfo{author}{\bibfnamefont{I.}~\bibnamefont{{Szapudi}}},
  \bibinfo{author}{\bibfnamefont{D.~J.} \bibnamefont{{Eisenstein}}},
  \bibinfo{author}{\bibfnamefont{N.}~\bibnamefont{{Afshordi}}},
  \bibinfo{author}{\bibfnamefont{T.}~\bibnamefont{{Budavari}}},
  \bibinfo{author}{\bibfnamefont{I.}~\bibnamefont{{Csabai}}},
  \bibinfo{author}{\bibfnamefont{J.~A.} \bibnamefont{{Frieman}}},
  \bibinfo{author}{\bibfnamefont{J.~E.} \bibnamefont{{Gunn}}},
  \bibinfo{author}{\bibfnamefont{D.}~\bibnamefont{{Johnson}}}, \emph{et~al.},
  \bibinfo{year}{2003}, \bibinfo{journal}{ArXiv Astrophysics e-prints}
  \eprint{arXiv:astro-ph/0307335}.

\bibitem[{\citenamefont{{Sheth} and {van de Weygaert}}(2004)}]{Sheth2004}
\bibinfo{author}{\bibnamefont{{Sheth}}, \bibfnamefont{R.~K.}}, and
  \bibinfo{author}{\bibfnamefont{R.}~\bibnamefont{{van de Weygaert}}},
  \bibinfo{year}{2004}, \bibinfo{journal}{\mnras}
  \textbf{\bibinfo{volume}{350}}, \bibinfo{pages}{517}.

\end{thebibliography}

\end{document}